\documentclass[ a4paper, 12pt ]
  {article}

\usepackage[dvips]{graphicx}

\newcommand{\NP}[1]{{\it Nucl.\ Phys.}\ {\bf #1}}
\newcommand{\ZP}[1]{{\it Z.\ Phys.}\ {\bf #1}}
\newcommand{\PL}[1]{{\it Phys.\ Lett.}\ {\bf #1}}
\newcommand{\PR}[1]{{\it Phys.\ Rev.}\ {\bf #1}}

\newcommand{\IJMP}[1]{{\it Int.\ J.\ Mod.\ Phys.}\ {\bf #1}}
\newcommand{\be}{\begin{equation}}
\newcommand{\ee}{\end{equation}}
\newcommand{\bea}{\begin{eqnarray}}
\newcommand{\eea}{\end{eqnarray}}
\newcommand{\bfk}{\mbox{$\mathbf {k}$}}
\newcommand{\bfq}{\mbox{$\mathbf {q}$}}
\newcommand{\pup}{p^\uparrow}
\newcommand{\pdown}{p^\downarrow}
\newcommand{\qup}{q^\uparrow}
\newcommand{\qdown}{q^\downarrow}
\newcommand{\bfp}{\mbox{$\mathbf p$}}
\newcommand{\bfP}{\mbox{$\mathbf P$}} 
\newcommand{\Lup}{\Lambda^\uparrow} 
\newcommand{\Aup}{A^\uparrow} 
\newcommand{\hup}{h^\uparrow} 
\newcommand{\hdown}{h^\downarrow} 

\begin{document}

\begin{center}
\bf{Sivers effect and transverse single spin \\ 
    asymmetries in Drell-Yan processes}\footnote{   
Talk delivered by U.~D'Alesio at the 
``15th International Spin Physics Symposium'', SPIN2002, September 9-14, 2002, 
Brookhaven National Laboratory, Upton (NY), USA. } 

\vspace*{0.6cm}

{\sf M.~Anselmino$^1$, U.~D'Alesio$^2$, F.~Murgia$^2$}
\vskip 0.5cm
{\it $^1$ Dipartimento di Fisica Teorica, Universit\`a di Torino and \\
          INFN, Sezione di Torino, Via P. Giuria 1, I-10125 Torino, Italy}\\
\vspace*{0.15cm}
{\it $^2$ Istituto Nazionale di Fisica Nucleare, 
Sezione di Cagliari \\
and Dipartimento di Fisica,  Universit\`a di Cagliari \\
C.P. 170, I-09042 Monserrato (CA), Italy} \\
\end{center}

\vspace*{0.5cm}

\begin{abstract}
Sivers asymmetry, adopted to explain 
transverse single spin asymmetries (SSA) observed in inclusive pion 
production, $\pup \, p \to \pi \, X$ and $\bar{p}^\uparrow \, p \to \pi \, X$, 
is used here 
to compute SSA in Drell-Yan processes; in this case, by considering the 
differential cross section in the lepton-pair invariant mass, rapidity and 
transverse momentum, other mechanisms which may originate SSA cannot 
contribute. Estimates for RHIC experiments are given.
\end{abstract}

\vspace*{0.5cm}

Single spin asymmetries in high energy inclusive processes are a unique
testing ground for QCD; they cannot originate from the simple spin pQCD 
dynamics -- dominated by helicity conservation -- but need some non
perturbative chiral-symmetry breaking in the large distance physics. 

Among the best known transverse single spin asymmetries (SSA) let us 
mention: 
$i)$ the large polarization of $\Lambda$'s and other 
hyperons produced in $p \, N \to \Lup \, X$; 
$ii)$ the large asymmetry 
$A_N = \frac{d\sigma^\uparrow - d\sigma^\downarrow}
           {d\sigma^\uparrow + d\sigma^\downarrow}
$
observed in $\pup \, p \to \pi \, X$ and  
$\bar p^\uparrow \, p \to \pi \, X$ processes; 
$iii)$ 
the similar azimuthal asymmetry observed in  
$\ell \, \pup \to \ell \, \pi \, X$. 

Several models \cite{prag} to explain the data within QCD 
dynamics can be found in the literature; 
here we focus on a phenomenological approach based on the 
generalization of the factorization theorem with the inclusion of parton 
intrinsic motion $\bfk_\perp$. 
The cross section for a generic process 
$A\,B \to C\,X$ then reads:
\be
d\sigma = \sum_{a,b,c} \hat f_{a/A}(x_a,\bfk_{\perp a}) \otimes 
\hat f_{b/B}(x_b, \bfk_{\perp b}) \otimes
d\hat\sigma^{ab \to c \dots}(x_a, x_b, \bfk_{\perp a}, \bfk_{\perp b}) 
\otimes \hat D_{C/c}(z, \bfk_{\perp C})\>,
\label{ltgen}
\ee
where the $\hat f$'s ($\hat D$'s) are the $\bfk_\perp$ dependent 
parton distributions (fragmentation functions).
  
Even if Eq.~(\ref{ltgen}) is not formally proven in general, 
it has been shown that
intrinsic $\bfk_\perp$'s are indeed necessary in order to be able to 
explain, within pQCD and the factorization scheme, data on (moderately) 
large $p_{_T}$
production of pions and photons \cite{pkt}.

When dealing with polarized processes the introduction of $\bfk_\perp$ 
dependences opens up the way to many possible spin effects; these can be 
summarized, at leading twist, by new polarized distribution functions 
and fragmentation functions,
\bea
\Delta^Nf_{q/\pup}  \!\!\! &\equiv& \!\!\!
\hat f_{q/\pup}(x, \bfk_{\perp})-\hat f_{q/\pdown}(x, \bfk_{\perp}) =  
\hat f_{q/\pup}(x, \bfk_{\perp})-\hat f_{q/\pup}(x, - \bfk_{\perp}) \>,
\label{delf1} \\
\Delta^Nf_{\qup/p}  \!\!\! &\equiv& \!\!\!
\hat f_{\qup/p}(x, \bfk_{\perp})-\hat f_{\qdown/p}(x, \bfk_{\perp}) =
\hat f_{\qup/p}(x, \bfk_{\perp})-\hat f_{\qup/p}(x, - \bfk_{\perp}) \>, 
\label{delf2}\\
\!\!\!\! \Delta^N D_{h/\qup} \!\!\!\! &\equiv& \!\!\!\!
\hat D_{h/\qup}(z, \bfk_{\perp}) - \hat D_{h/\qdown}(z, \bfk_{\perp}) =
\hat D_{h/\qup}(z, \bfk_{\perp})-\hat D_{h/\qup}(z, - \bfk_{\perp}) \>,
\label{deld1} \\
\!\!\!\! \Delta^N D_{\hup/q} \!\!\!\! &\equiv& \!\!\!\!
\hat D_{\hup/q}(z, \bfk_{\perp}) - \hat D_{\hdown/q}(z, \bfk_{\perp}) =
\hat D_{\hup/q}(z, \bfk_{\perp})-\hat D_{\hup/q}(z, - \bfk_{\perp}) \>,
\label{deld2} 
\eea
which have a clear meaning if one pays attention to the arrows denoting the
polarized particles. All the above functions vanish when $k_\perp=0$ and 
are na\"{\i}vely $T$-odd. The ones in Eqs. (\ref{delf2}) and (\ref{deld1}) 
are chiral-odd, while the other two are chiral-even.   
The fragmentation in Eq. (\ref{deld1}) is the Collins function \cite{col},
while the distribution in Eq. (\ref{delf1}) was first introduced by Sivers 
\cite{siv}. 
Some of the above functions have been widely used for a phenomenological 
description of the observed SSA \cite{noi}.  

Despite its successful phenomenology, the Sivers function was always
a matter of discussions and its very existence rather controversial; in fact
in Ref.~\cite{col} a proof of its vanishing was given, based on time-reversal 
invariance. 
Ways out based on initial state interactions or 
non standard time-reversal properties \cite{dra} were discussed. 
Very recently a series of papers \cite{newsiv} 
have resurrected Sivers asymmetry 
in its full rights: a quark-diquark model
calculation has given an explanation of 
the HERMES azimuthal asymmetry different from the  
Collins effect and has shown 
that initial state interactions can give rise to SSA in Drell-Yan
processes. 
Moreover Collins recognized 
that $i)$ such a new mechanism is compatible with factorization and is due  
to the Sivers asymmetry (\ref{delf1}), $ii)$ his original proof 
of the vanishing of $\Delta^Nf_{q/\pup}$ is incorrect. 

Some issues concerning factorizability and universality of these
effects are still open to debate; however, we feel now confident to use 
Sivers effects -- and equally all functions in 
Eqs. (\ref{delf1})-(\ref{deld2}) -- in SSA phenomenology. 
The natural process to test the Sivers asymmetry is Drell-Yan where  
there cannot be any effect in fragmentation processes and, 
by suitably integrating over some final configurations, 
other possible effects vanish. 
SSA in Drell-Yan processes are particularly important now, as 
ongoing or imminent experiments at RHIC will be able to measure them. 

Let us consider a Drell-Yan process, that is the production of 
$\ell^+\ell^-$ pairs in the collision of two hadrons $A$ and $B$: 
the difference between 
the single transverse spin dependent cross sections $d\sigma^\uparrow$ for 
$A^\uparrow \, B \to \ell^+ \, \ell^- \, X$ and $d\sigma^\downarrow$ for
$A^\downarrow \, B \to \ell^+ \, \ell^- \, X$ 
from the Sivers asymmetry of Eq. (\ref{delf1}), 
is
\be
d\sigma^\uparrow - d\sigma^\downarrow = 
\sum_{ab} \int \left[ dx_a \, d^2\bfk_{\perp a} 
\, dx_b \, d^2\bfk_{\perp b} \right] \, 
\Delta^Nf_{a/A^\uparrow}(x_a,\bfk_{\perp a}) \,
\hat f_{b/B}(x_b,\bfk_{\perp b}) \,
d\hat\sigma^{ab \to \ell^+\ell^-}
\!.
\label{ddy1}
\ee

We consider the differential cross section in the variables 
$ M^2 = (p_a + p_b)^2 $, $ y $ and $ \bfq_{_T} $, 
that is the squared invariant mass, the rapidity and the transverse momentum 
of the lepton pair. 
Notice that we do not look at the 
angular distribution of the lepton pair production plane, which is 
integrated over.  

We take the hadron $A$ as moving along the positive $z$-axis, 
in the $A$-$B$ c.m. frame and measure the transverse polarization 
of hadron $A$, $\bfP_{\!A}$, along the $y$-axis. 

In the kinematical regions such that: 
$q_{_T}^2 \ll M^2 \ll M_Z^2$ and $k_{\perp a,b}^2 \simeq q_{_T}^2$, 
the asymmetry becomes
\be
A_N = \frac
{\sum_q e_q^2 \int d^2\bfk_{\perp q} \, d^2\bfk_{\perp \bar q} \>
\delta^2(\bfk_{\perp q} + \bfk_{\perp \bar q} - \bfq_{_T}) \>
\Delta^Nf_{q/\Aup}(x_q, \bfk_{\perp q}) \>
\hat f_{\bar q/B}(x_{\bar q}, \bfk_{\perp \bar q})}
{2 \sum_q e_q^2 \int d^2\bfk_{\perp q} \, d^2\bfk_{\perp \bar q} \>
\delta^2(\bfk_{\perp q} + \bfk_{\perp \bar q} - \bfq_{_T}) \>
\hat f_{q/A}(x_q, \bfk_{\perp q}) \>
\hat f_{\bar q/B}(x_{\bar q}, \bfk_{\perp \bar q})} \>, \label{ann}
\ee
where $x_q\simeq\frac{M}{\sqrt s} \, e^y $ 
and $x_{\bar q}\simeq\frac{M}{\sqrt s} \, e^{-y}$, with $a,b = q, \bar q$  
and $q = u, \bar u, d, \bar d, s, \bar s$.    

On the other hand the SSA generated by the distribution 
function in Eq. (\ref{delf2})  
would lead to a con\-tri\-bu\-tion of the kind \cite{dan}   
\be
\sum_q h_{1q}(x_q,\bfk_{\perp q}) \otimes 
\Delta^N f_{\bar q^\uparrow/B}(x_{\bar q}, \bfk_{\perp \bar q}) \otimes 
d\Delta \hat\sigma^{q\bar q  \to \ell^+\ell^-} \>, \label{andan}
\ee
where $h_{1q}$ is the transversity of quark $q$ (inside hadron $A$) 
and $d\Delta \hat\sigma$ is the double transverse spin asymmetry 
$d\hat\sigma^{\uparrow\uparrow} - d\hat\sigma^{\uparrow\downarrow}$.
Such an elementary asymmetry has a $\cos2\phi$ dependence \cite{dan}, 
where $\phi$ is the angle between the transverse polarization direction
and the normal to the $\ell^+\ell^-$ plane; when integrating 
over all final angular distributions of the $\ell^+\ell^-$ pair -- as we do 
-- the contribution of Eq. (\ref{andan}) vanishes. 
 
Analogously, other mechanisms \cite{htw}, based on higher twist quark-gluon 
correlation functions,  
lead to expressions of $A_N$ vanishing upon integration
over the leptonic angles. 

In order to give numerical estimates, we introduce here 
a simple model for the Sivers asymmetry (\ref{delf1}), and for the
unpolarized distributions, which is similar to the one introduced for 
the polarizing fragmentation function (see Eq.~\ref{deld2}) 
in Ref.~\cite{noi2}.

Let us start from the most general expression for the number density  
of unpolarized 
quarks $q$, inside a proton with transverse polarization $\bfP$ and 
three-momentum $\bfp$. One has 
\be
\hat f_{q/\pup}(x, \bfk_{\perp}) = \hat f_{q/p}(x, k_{\perp})
+ \frac{1}{2} \, \Delta^N f_{q/\pup}(x, k_{\perp}) \> \hat{\bfP} \cdot
\hat{\bfp} \times \hat{\bfk}_{\perp}\>. \label{polden}
\ee
In our configuration one simply has
$\hat{\bfP} \cdot \hat{\bfp} \times \hat{\bfk}_\perp =
(\hat{\bfk}_\perp)_x=\cos\phi_{k_\perp}$.

We consider simple factorized and Gaussian forms 
(see also \cite{adm02}):
\be
\hat f_{q/p}(x, k_\perp) = f_{q/p}(x)\,g(k_\perp) =
f_{q/p}(x) \,\frac{\beta^2}{\pi}\, e^{-\beta^2 \, k_\perp^2} \>; 
\label{modu}
\ee
by imposing the positivity bound 
$ |\Delta^N f_{q/\pup}(x, k_{\perp})| \leq 
 2 \,\hat f_{q/p}(x, k_{\perp})   $, we can write
\be
\Delta^Nf_{q/\pup}(x, k_\perp) = 2 \,  {\mathcal N}_q(x) \,
f_{q/p}(x) \, \frac{\beta^2}{\pi} \, \sqrt{2\,e\,(\alpha^2 - \beta^2)}\,
k_\perp \, e^{-\alpha^2 k_{\perp}^2}  \label{del2}\>,
\ee
with
\be
{\mathcal N}_q(x) = N_q\,x^{a_q}(1-x)^{b_q}\,
\frac{(a_q+b_q)^{(a_q+b_q)}}{a_q^{a_q}\,b_q^{b_q}}\,,\quad
|N_q|\leq 1\,\>.
\label{nqx}
\ee
Inserting the above choice of $\Delta^Nf(x, \bfk_\perp)$ and 
$\hat f(x, \bfk_\perp)$ into Eq. (\ref{ann}) one can perform analytical
integrations; assuming $\beta$ independent of $x$ (see below) one gets 
\bea
A_N(M,y,\bfq_{_T}) &=&
{\mathcal Q}(q_{_T},\phi_{q_{_T}})\>{\mathcal A}(M,y) \nonumber\\
&=&2\,\frac{r^2}{(1+r)^2}\>\left(\,2\,e\,\frac{1-r}{r}\,\right)^{1/2}\>
\beta \,q_{_T} \,\cos\phi_{q_{_T}}\>\exp\,\left[\,-\frac{1}{2}\,\frac{1-r}{1+r}\,
\beta^2\,q_{_T}^2\,\right]\nonumber\\
&\times&\>\frac{1}{2}\>\frac{\sum_q e_q^2 \, \Delta^Nf_{q/\pup}(x_q) \,
f_{\bar q/p}(x_{\bar q})} {\sum_q e_q^2 \, f_{q/p}(x_q) \, 
f_{\bar q/p}(x_{\bar q})} \> \cdot
\label{anr}
\eea
where $\phi_{q_{_T}}$ is the azimuthal angle of $\bfq_{_T}$ 
and $r\equiv\beta^2 / \alpha^2<1$. ${\mathcal Q}(q_{_T})$ has a maximum when 
$q_{_T} = q_{_T}^M = \sqrt{(1+r)/(1-r)}/\beta $,  
where its value is ${\mathcal Q}(q_{_T}^M)\equiv {\mathcal Q}_M = 
[\,2\,r/(1+r)\,]^{3/2}$. Notice that only the position of the maximum
depends on the parameter $\beta$.

All the parameters of the model 
have been fixed in a complementary analysis  of 
unpolarized inclusive particle production  and pion SSA \cite{dm02}.  
As a result of this study we have: 
$\beta=1.25\,({\rm GeV}/c)^{-1}$ 
($\langle\,k_\perp^2\,\rangle^{1/2} = 0.8$ GeV/$c$, independent of $x$), 
and 
\bea
N_u &=&  \phantom{-}0.5 \qquad a_u = 2.0 \qquad b_u = 0.3 \>,\nonumber\\
N_d &=& -1.0 \qquad a_d = 1.5 \qquad b_d = 0.2\>, \qquad 
\qquad\quad r\simeq 0.7\>.
\label{nab}
\eea

For the unpolarized partonic distributions, $f_{q/p}(x)$,  
we adopt the GRV94 set \cite{grv94}.

One further uncertainty concerns the sign of the asymmetry: as noticed 
by Collins and checked by Brodsky \cite{newsiv}, the Sivers asymmetry 
has opposite signs in Drell-Yan and SIDIS, respectively related 
to $s$-channel and $t$-channel elementary reactions. As in $p-p$ 
interactions we expect that large $x_F$ regions are dominated by 
$t$-channel quark processes, we think that the Sivers function
extracted from $p-p$ data should be opposite to that contributing
to D-Y processes. Our numerical estimates will then be given with the same
parameters as in Eq.~(\ref{nab}), {\it changing the signs of} $N_u$ and $N_d$. 
Given these considerations, even a simple comparison of the sign of our
estimates with data might be significant.

\begin{figure}
 \hspace*{0.1cm}
 \includegraphics[angle=-90, width=.5\textwidth]{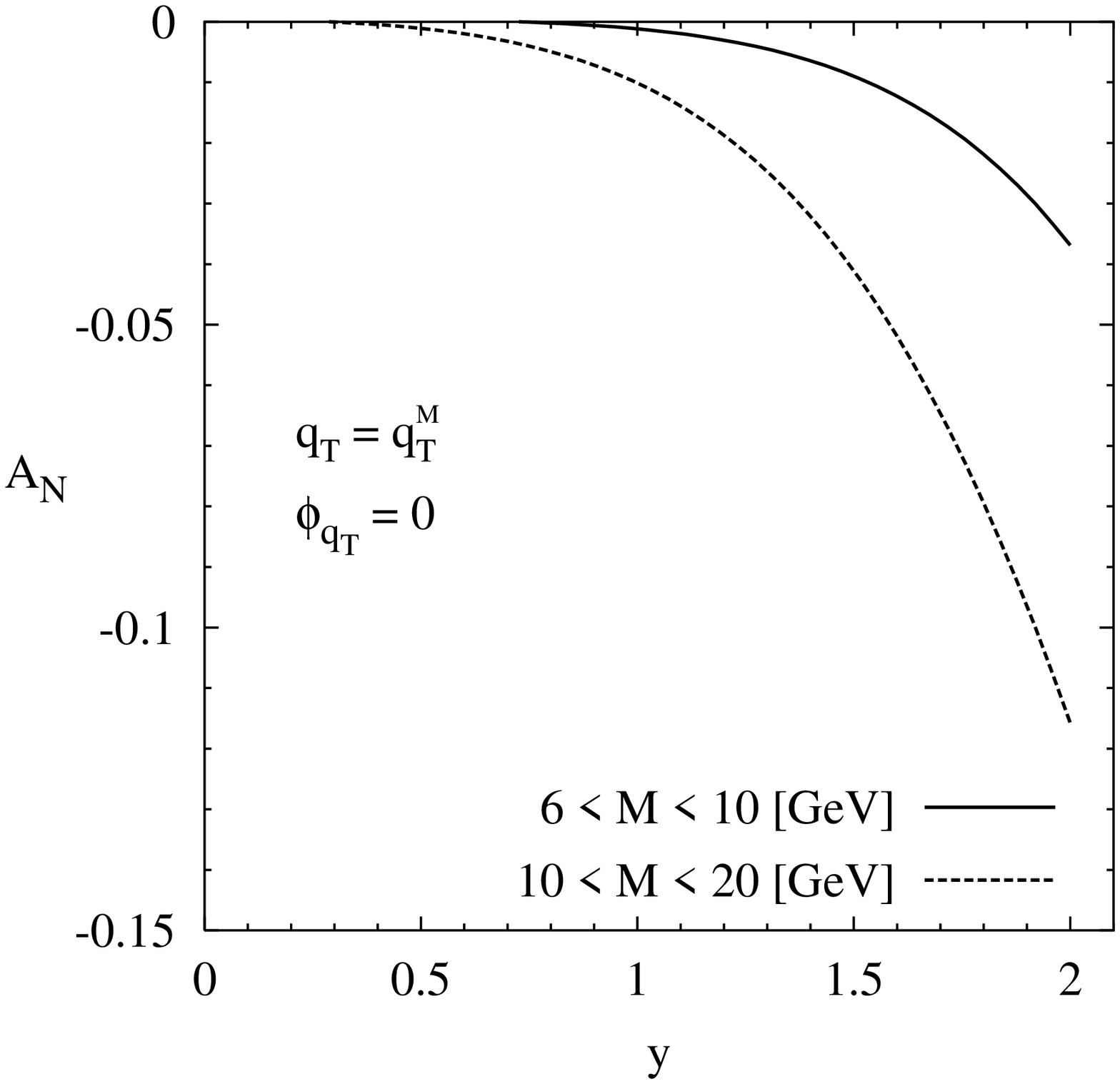}
 \includegraphics[angle=-90, width=.5\textwidth]{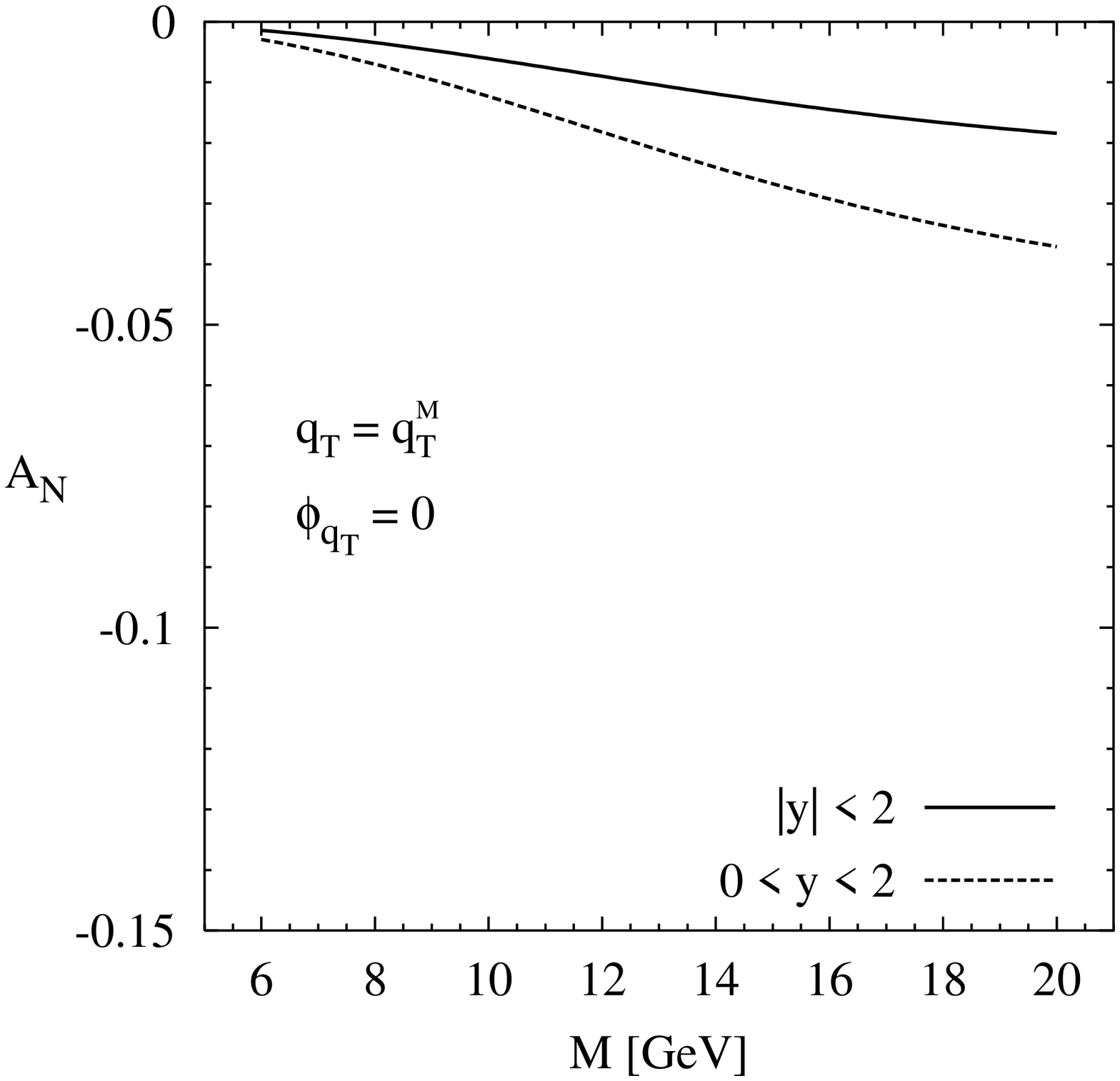}
 \caption{\small Single spin asymmetry $A_N$ for the Drell-Yan process,
 at RHIC energies, $\sqrt{s}$~=~200 GeV, as a function of $y$ and
 averaged over $M$ (left) and as a function of
 $M$ and averaged over $y$ (right), (see text).}
\label{fig}
\end{figure}

In Fig.~\ref{fig} we show $A_N$ at $\sqrt s = 200$ GeV as a function of $y$
averaged over two kinematical ranges $6\leq M \leq 10$ GeV and 
$10 \leq M \leq 20$ GeV (on the left) and 
as a function of $M$ averaged over the ranges $|y|<2$
and $0<y<2$ (on the right). 
We have fixed $q_{_T}=q_{_T}^M$ ($\simeq 1.9$ GeV/$c$), and 
$\phi_{q_{_T}}=0$, which maximizes the $\bfq_{_T}$-dependent part of the 
asymmetry; on the other hand $A_N$ is reduced by a factor of 50\%  
at $q_{_T}\simeq 0.6$ GeV/$c$.

We can also consider the asymmetry averaged
over $\bfq_{_T}$ up to a value of $q_{_T} = q_{_{T1}}$ 
(integrating over $\phi_{q_{_T}}$ in the range $[0,\pi/2]$ only,
otherwise one would get zero). 
In our simple model (for a full account of this study see
\cite{adm02}),  
for $q_{_{T1}} \ge 1.7 $ GeV/$c$ we would get $\langle A_N \rangle\simeq 0.4
\,A_N(q_{_T}^M)$ (for $q_{_{T1}} = 0.6 $ GeV/$c$ $\langle A_N \rangle\simeq 0.2
\,A_N(q_{_T}^M)$ ).

Our numerical estimates show that $A_N$ can be well measurable within 
RHIC expected statistical accuracy. The actual values depend on the
assumed functional form of the Sivers function and its role with valence 
quarks only. 

Transverse single spin phenomenology, within QCD dynamics and the
factorization scheme, is a rich and interesting subject. 
It combines 
simple pQCD spin dynamics with new long distance properties of quark
distribution and fragmentation; the experimental measurements are relatively
easy and clear, many have been and many more will be performed in the 
near future, both at nucleon-nucleon and lepton-nucleon facilities. 

Very recently a large transverse SSA (contrary to naive expectations) 
has been observed at $\sqrt s =$ 200 GeV (at RHIC) 
in $\pup \, p \to \pi \, X$ processes \cite{bnl}; a reasonable
agreement with these preliminary data has been found in our approach.

We have presented here the explicit formalism for computing single transverse 
spin asymmetries in Drell-Yan processes, within a generalized QCD 
factorization theorem formulated with $\bfk_\perp$ dependent distribution
functions. Simple Gaussian forms have been assumed and available data from 
other processes have been exploited, in order to give estimates for single
spin effects in D-Y production at RHIC, which should be of interest for the 
incoming measurements. Again, sizable and measurable values have been found.  

\vspace*{0.5cm}

\noindent
{\bf Acknowledgments} 

\vspace*{0.2cm}

One of us (U.D.) would like to thank the organizers for their kind
invitation to a fruitful and interesting Symposium. U.D. 
and F.M. thank COFINANZIAMENTO MURST-PRIN for partial support. 

{\small

}
\end{document}